# Case Study: Horizontal Side-Channel Analysis Attack against Elliptic Curve Scalar Multiplication Accelerator under Laser Illumination

Dmytro Petryk[1], Ievgen Kabin[1], Peter Langendoerfer[1,2] and Zoya Dyka[1,2]
[1] *IHP – Leibniz-Institut für innovative Mikroelektronik,* Frankfurt (Oder), Germany
[2] *BTU Cottbus-Senftenberg,* Cottbus, Germany
{petryk, kabin, langendoerfer, dyka}@ihp-microelectronics.com

*Abstract* — Devices employing cryptographic approaches have to be resistant to physical attacks. Side-Channel Analysis (SCA) and Fault Injection (FI) attacks are frequently used to reveal cryptographic keys. In this paper, we present a combined SCA and laser illumination attack against an Elliptic Curve Scalar Multiplication accelerator using a differential probe from Teledyne LeCroy. Our experiments show that laser illumination increases the power consumption of the chip, especially its static power consumption but the success of the horizontal power analysis attacks was changed insignificantly. We assume that using a laser with a high laser beam power and concentrating on measuring and analysing only static current can improve the attack success significantly. The horizontal attacks against public key cryptosystems exploiting the Static Consumption under Laser Illumination (SCuLI attacks) are novel and their potential is not investigated yet. These attacks can be especially dangerous against cryptographic chips manufactured in scaled technologies. If such attacks are feasible, appropriate countermeasures have to be proposed in the future.

*Keywords* — *Security, Side-Channel Analysis (SCA), laser illumination, power consumption, Static Consumption under Laser Illumination (SCuLI) attacks.*

## I. INTRODUCTION

Physical attacks pose a great threat for today's semiconductor devices, in which cryptographic approaches are frequently used to ensure security requirements such as confidentiality, data integrity, availability of services, and authentication. The strength of cryptographic approaches is based on the secrecy of the key(s) used, where their length depends on the applied algorithm and security requirements. The state-of-the-art approaches provide a proved level of security using keys of recommended lengths, i.e. the algorithms cannot be compromised by cryptanalysis nor brute-force attacks in a reasonable time. The issue is that in real world scenarios the devices are usually physically accessible, can be stolen and attacked in a lab. Side-Channel Analysis (SCA) as well as Fault Injection (FI) attacks are frequently used to breach the device's protection. SCA is based on the analysis of the device's physical emissions. Analysis of the measured emissions can be done using many traces (vertical attack) or using a single measured trace (horizontal attack).

FI attacks aim at manipulating the device's normal operation by inducing a fault while the device is in operation. Attacks using lasers are frequently used due to their localized area of influence and accurate timing. The minimal laser spot size is limited due to diffraction so that laser-based attacks are frequently performed against chips manufactured in "old" technologies, i.e. technologies with large node size. E.g, in [1], it was feasible to manipulate a single transistor, because the illuminated microcontroller was manufactured in a 1.2 μm technology while the laser beam spot size was 1 μm. As technology advances, precision of laser-based attacks is decreasing due to the fact that many cells are illuminated simultaneously even when applying state-of-the-art laser with diffraction-limited spot size. At the same time technology downscaling offers reduced power consumption that means the circuit can be more vulnerable to laser illumination (LI), i.e. the gap between no fault and permanent fault can be so small that untraceable transient faults can be infeasible. Due to these facts, precise single bit faults are expected to be unfeasible in future. We expect that, in future, the main focus will be on illuminating a critical block with the goal to improve SCA by increasing its power consumption without injecting a fault. In such an attack the contribution of the illuminated block in the measured power trace will be "more visible", i.e. it can improve extraction of the secret key processed.

In this paper, we present the evolution of laser-based attacks from FI towards their future evolution into SCA under LI attacks against IHP's Elliptic Curve Scalar Multiplication accelerator. We implemented a single trace – horizontal – attack under laser illumination while performing a scalar multiplication that involves a private key. Opposite to attacks published in the past we try to increase the area illuminated by the laser to estimate the feasibility of such attacks using a single laser.

This paper is structured as follows. Section II gives a brief introduction to SCA and FI attacks as well as an overview of works applying LI to improve SCA attacks. Section III describes our Elliptic Curve Scalar Multiplication accelerator, our setup, its configuration and settings. Section IV represents and discusses the attack results. Section V concludes this work.

## II. COMBINING SCA WITH LASER ILLUMINATION

### A. SCA and Laser illumination attacks

*SCA attacks* are feasible due to changing physical parameters while the device performs operations, e.g. power consumption, electromagnetic emanation, time to perform operations, etc. Analysing measured parameters using for example statistical analysis methods can reveal a secret/private key processed during the observed cryptographic operation. To counter SCA attacks many algorithmic approaches are known. Additionally, noise can hide the contribution of the security-critical block(s), whereby the activity of the block(s) functioning in parallel to the security-critical block is a kind of noise.

*Laser illumination attacks* are feasible due to the known interaction of light with semiconductors. Using laser illumination, it is feasible (1) to inject a fault in a logic cell via switching one of the illuminated transistors, or (2) to increase the power consumption of the illuminated logic cells without switching transistor(s) of the attacked circuit.

Increasing the contribution of the power consumption of a security-critical block via its illumination allows to increase the success of the key extraction, i.e. attacks can be improved combining SCA and Laser Illumination without fault injections.

### B. State-of-the-Art

Attacks exploiting measurements of a side-channel parameter under laser illumination, without introducing any faults, are rare in the literature. In the past, only few works reported SCA attacks with laser illumination. Authors in [2] used a laser beam to increase the power consumption of a circuit by illuminating an SBOX block during the last round of DES implemented in an FPGA. SBOX blocks are the security-critical blocks in symmetric cryptographic approaches. As a result, the authors were able to perform successful differential power analysis attacks with a reduced number of power traces as well as to recover a sub-key, which was unfeasible without laser illumination. In [2], the dynamic power consumption was measured and analysed. In [3] the author used laser illumination to detect access events of SRAM memory. In [4], the authors used laser illumination to extract data stored in EEPROM. While illuminating the EEPROM sense amplifiers, it was possible to retrieve the program stored in the memory. In [5], the static power consumption of the chip was measured under laser illumination without injecting a fault. According to the obtained results the laser illumination increased the static power consumption significantly. Moreover, the increased static current is data dependent, i.e. the static power consumption of an illuminated logic cell depends on its inputs. In this work, we focus on the measurement of dynamic power consumption of a hardware accelerator for an asymmetric cryptographic approach under laser illumination.

### III. CRYPTOGRAPHIC ACCELERATOR AND ATTACK SETUP

The attacked accelerator was manufactured at IHP [6] in its 250 nm technology. It is designed to perform Elliptic Curve (EC) point multiplication with a scalar, denoted as *kP* operation, where *P* is a point on an EC and *k* is a secret long binary number, for example a private key. The *kP* operation is implemented according to a slightly modified Montgomery ladder (see Algorithm A2 in [7] using Lopez-Dahab projective coordinates for EC point representation, for the standardised NIST EC *B-233* [8]. Additional details about the IHP's accelerator can be found in [9].

To perform an SCA attack under LI, the equipment available at IHP was used. The setup is shown schematically in **Fig. 1**.

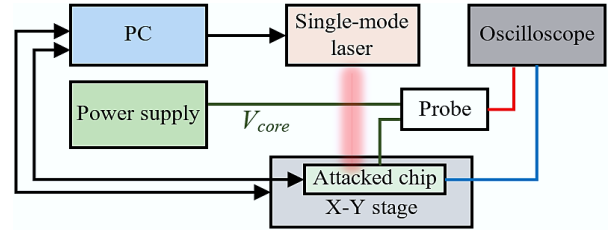

**Fig. 1.** Schematic representation of the attack setup.

The applied measurement setup consists of: a PC used to communicate with the attacked cryptographic accelerator, commercially available differential probe from Teledyne LeCroy [10], a stable power supply, and a modified laser station from Riscure [11] with a red (808 nm) single-mode Pulse-on-Demand Module from Alphanov [12]. The probe was connected to $V_{core}$ and an oscilloscope. Details about the laser equipment as well as the evaluation of its parameters can be found in [13].

To perform the attacks the manufactured cryptographic accelerator was bonded to printed circuit board (PCB) without any packaging. The PCB was fixed on the metal board that was put onto a high-precision X-Y stage. The X-Y stage is controlled using the Riscure Inspector software.

During our experiments we tracked the correctness of calculations performed by the accelerator with and without laser illumination, i.e. if there are no disruptions (no injected faults), while the chip performs the *kP* operation. The chip operating frequency was set to 4 MHz. To start measurements at the correct time, one of the chip's pins was used as a trigger since the signal at the pin is in logic state high during the execution of the *kP* operation. All oscilloscope measurements in this work were performed with a sampling rate of 5 GS/s resulting in 1250 samples per clock cycle.

The *kP* accelerator was illuminated through the front-side. No decapsulation of the chip was required. We used a single-mode laser due to the known power distribution (Gaussian) and its ability to operate in a Continuous Wave (CW) mode. The last prerequisite is important since the *kP* operation takes ~3.2 ms and the laser should be able to generate a uniform beam with a constant output power during the execution of the *kP* operation[1]. In this work, we describe the configuration of the setup as well as parameters used to perform the experiments in detail with the goal to allow reproducibility of the experiments done. The issue of experiments reproducibility was covered in [14].

The single-mode laser was controlled by Alphanov control software. The laser beam output power is controlled by the current[2] set in the control software, it is represented in mA, the maximum current per channel is limited at 450.0 mA and corresponds to 100 % of the laser power in the CW mode. The laser has two channels (PDM 2+) that generate a beam. In our attacks, we used both channels simultaneously. The selected area was illuminated for ~5 s before we started the execution of the *kP* operation under laser illumination. We used a long working distance NIR 5× objective from Mitutoyo [16] with the goal to have the biggest laser beam spot at focus in the setup. To further increase the illuminated area, in some experiments, we moved away from the focus to a point where the area illuminated by laser is ~30 times larger than the one in the focus.

---

[1] Lasers used to perform optical FI attacks are designed to operate in a pulsed mode, i.e. in a range up to hundreds of μs [15].

[2] The control current in Alphanov control software is denoted as a DC or Offset parameter.

## IV. Performed attacks and their Results

For the analysis of the measured traces we used the IHP SCA tool described in [9]. We compressed the measured trace using sum of squared values method and applied the comparison-to-the-mean method to extract 54 key candidates. We used the relative correctness of each key candidate to evaluate the attack success. The relative correctness shows in percent the number of correctly revealed bits for each key candidate. The relative correctness of 100% means that all bits of the key candidate are revealed correctly, i.e. the key candidate is identical to the real scalar *k* processed during the executed *kP* operation.

### A. SCA without laser illumination

First, we measured a reference trace, i.e. a trace without laser illumination, using the differential probe from Teledyne LeCroy [10]. The trace is shown in **Fig. 2**-*(a)*, and the results of its analysis are shown in **Fig. 2**-*(b)*.

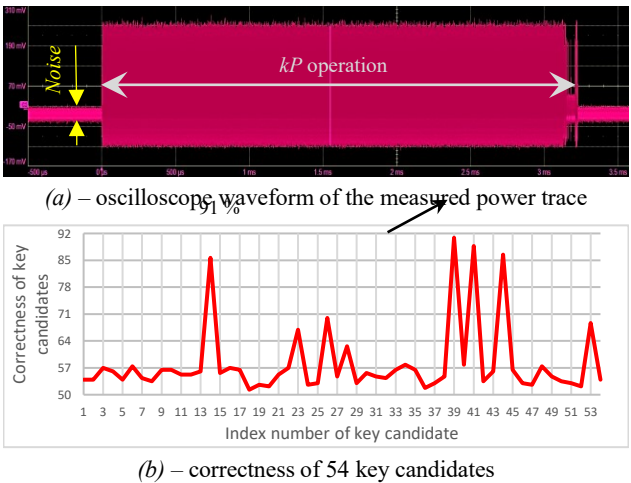

(a) – oscilloscope waveform of the measured power trace

(b) – correctness of 54 key candidates

**Fig. 2.** Results of SCA attack (without laser illumination).

Please note that processing a key bit requires only 54 clock cycles and is always performed using the same operation sequence. According to our analysis using the IHP SCA tool, the correctness of the best key candidate is 91 %, i.e. 211 bits of the 232 bits long scalar *k* were revealed.

### B. SCA under laser illumination

We illuminated with a laser the attacked EC cryptographic accelerator during the *kP* operation varying the laser beam parameters. Our goal was to clarify if the influence of the laser illumination on the power consumption can be observed and if the laser illumination influences the attack success.

The attacked chip has different blocks: field multiplier, field adder, registers, controller, and a multiplexer. It was expected that illuminating the field multiplier deteriorates the key extraction because this block is not an SCA leakage source. If the power consumption of the multiplier will be increased via laser illumination, its contribution to the total power consumption will hide, at least partially, the contribution of other design blocks, i.e. the attack success will be decreased.

All our experiments described here were performed at a fixed position of the laser over the chip surface illuminating an area of the field multiplier. We captured 5 power traces under laser illumination. In all experiments with the laser we used the 5× objective. To illuminate a relatively big area we unfocused the laser beam. We used both channels of the single-mode laser.

TABLE I shows the power of the laser beam, the diameter of the laser beam spot and the attack success for all experiments. The laser beam spot size and the laser beam power in experiments 4, 5, 6 were selected to keep the same laser beam power per unit area (i.e. the same intensity). The spot sizes given in TABLE I are measured applying Full Width at Half Maximum (FWHM) measurement standard[3] using a laser beam profiler from Kokyo [17]. The last column in TABLE I shows the success rate of the attacks using the correctness of the best key candidate for each analysed trace.

TABLE I. LASER BEAM SETTINGS AND SUCCESS RATE OF PERFORMED HORIZONTAL ATTACK.

| Nr. of experiment | Laser beam power, % | Laser beam diameter $d$, μm | Correctness of the best key candidate δ, % |
|---|---|---|---|
| 1 | Reference trace (without illumination) | | 91 |
| 2 | 3 | 14 | 89 |
| 3 | 100 | 14 | 92 |
| 4 | 13 | 27 | 92 |
| 5 | 59 | 58 | 90 |
| 6 | 100 | 75 | 90 |

Please note that the differential probe measures not only the alternating component but also direct component of the voltage, i.e. the measured traces demonstrate the influence of the laser illumination on the dynamic as well as on the static power consumption of the attacked chip. **Fig. 3** shows a part of the measured power traces. Increasing the laser beam output power increases the power consumption of a chip. Our measurements also show that the laser beam power is the main parameter influencing the power consumption of the attacked chip: 100% of the laser beam illuminating the smallest area and the same energy illuminating 30 times larger area cause similar increase of power consumption of the illuminated chip, see black dashed and orange lines in **Fig. 3**. This means that an attacker can illuminate either a whole critical block or just a small part of it (but at higher laser beam power) causing a similar increase in the chip's power consumption.

---

[3] The laser beam spot sizes were measured using FWHM standard due to the fact that the laser beam spot sizes starting from a distance of 2520±12 μm from a focus is larger than the field of view of the used laser beam profiler, i.e. the laser beam spots cannot be (fully) captured by the laser beam profiler.

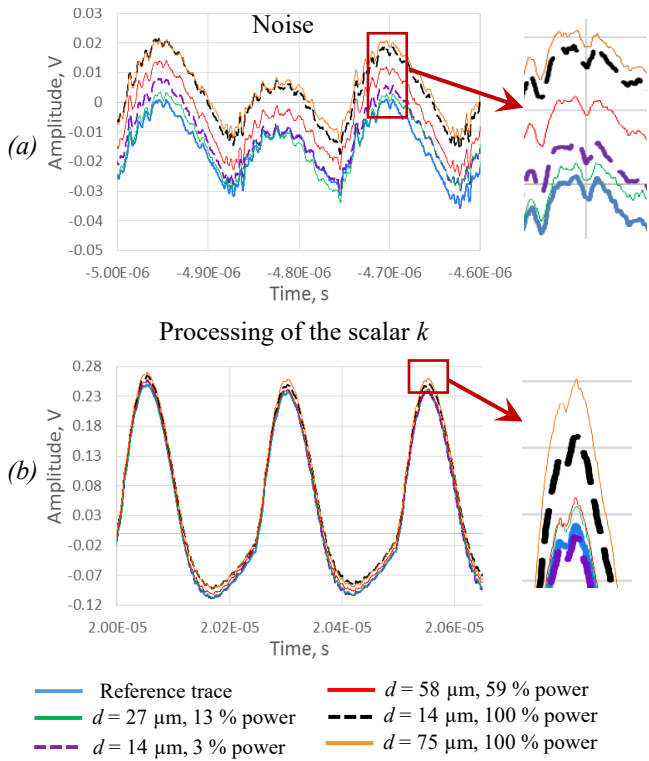

**Fig. 3.** A part of the measured power traces.

In our experiments, laser illumination changed the success of the attack insignificantly, i.e. the attack success is increased by 1 % and is decreased by 1-2 %, compared to the reference trace, see TABLE I experiments 3, 4 and 2, 5, 6, respectively. The attacked chip was manufactured with metal fillers above the transistors. These act as obstacles for the laser beam and can cause reflection, scattering, and absorption of photons. Additionally, we cannot exclude the possibility that other design blocks, and not only the field multiplier, were illuminated. Moreover, in 4 out of 5 experiments, the difference between the best key candidates is only 1 %, this can be caused by measurement tolerance influencing the analysis results. For example, analysing 3 reference traces measured on different days using the same differential probe, we obtained different attack success of 89.6 %, 90.9 % and 89.1 %. We expect that changes in power consumption of the attacked chip will be more noticeable applying a laser with higher output power and illuminating a larger area.

In future work, we plan to experiment with other laser sources with larger laser beam spots and with higher output power to improve horizontal SCA attacks combined with laser illumination. We expect that increasing the laser beam output power will cause bigger changes of the dynamic power consumption and a significant increase of the static power consumption of the attacked chip. Horizontal attacks against public key cryptosystems exploiting the Static Consumption under Laser Illumination (SCuLI attacks) are novel, and not investigated yet. Feasibility and potential of SCuLI attacks have to be evaluated. If they are feasible appropriate countermeasures have to be developed. We also plan to perform SCuLI attacks against chips manufactured in scaled technologies.

## V. Conclusion

In this work, we performed horizontal SCA against an Elliptic Curve Scalar Multiplication accelerator measuring its power traces under laser illumination. Our experiments clearly demonstrate the influence of the laser illumination on the power consumption of the attacked chip using the differential probe from Lecroy. Without laser illumination, the correctness of the best key candidate is 90 %. Analysis of the traces measured under laser illumination shows only small impact on the attack success. However, our measurements demonstrate that laser illumination influences the power consumption of the illuminated chip, especially of the static "component". The potential of the attacks exploiting the Static Consumption under Laser Illumination (SCuLI attacks) is not investigated yet. These attacks can be especially dangerous against cryptographic chips manufactured in scaled technologies.